\documentclass{elsart}

\input{epsf}

\newcommand{\Gam}{\mbox{$\textstyle\Gamma$}}
\newcommand{\kpi}{\mbox{$D^0\rightarrow K^-\pi^+$}}
\newcommand{\kk}{\mbox{$D^0\rightarrow K^+K^-$}}
\newcommand{\pipi}{\mbox{$D^0\rightarrow \pi^+\pi^-$}}
\newcommand{\kkovkpi}{\mbox{${\Gam{\textstyle (}\kk{\textstyle )}}\over
                     {\Gam{\textstyle (}\kpi{\textstyle )}}$}}
\newcommand{\pipiovkpi}{\mbox{${\Gam{\textstyle (}\pipi{\textstyle )}}\over
                       {\Gam{\textstyle (}\kpi{\textstyle )}}$}}
\newcommand{\kkovpipi}{\mbox{${\Gam{\textstyle (}\kk{\textstyle )}}\over
                      {\Gam{\textstyle (}\pipi{\textstyle )}}$}}

\begin{document}

\begin{frontmatter}
 
\title{
       Branching Fractions for $D^0 \rightarrow K^+K^-$
       and $D^0 \rightarrow \pi^+ \pi^-$, and a Search for {\em CP} 
       Violation in $D^0$ Decays\thanksref{subm}
      }
       \thanks[subm]{To be published in Physics Letters {\bf B}.}

\collab {Fermilab E791 Collaboration}

\author[umis]{E.~M.~Aitala},
\author[cbpf]{S.~Amato},
\author[cbpf]{J.~C.~Anjos},
\author[fnal]{J.~A.~Appel},
\author[taun]{D.~Ashery},
\author[fnal]{S.~Banerjee},
\author[cbpf]{I.~Bediaga},
\author[umas]{G.~Blaylock},
\author[stev]{S.~B.~Bracker},
\author[stan]{P.~R.~Burchat},
\author[ilit]{R.~A.~Burnstein},
\author[fnal]{T.~Carter},
\author[cbpf]{H.~S.~Carvalho},
\author[usca]{N.~J.~Copty},
\author[umis]{L.~M.~Cremaldi},
\author[yale]{C.~Darling},
\author[fnal]{K.~Denisenko},
\author[pueb]{A.~Fernandez},
\author[usca]{G.~Fox},
\author[ucsc]{P.~Gagnon},
\author[cbpf]{C.~Gobel},
\author[umis]{K.~Gounder},
\author[fnal]{A.~M.~Halling},
\author[cine]{G.~Herrera},
\author[taun]{G.~Hurvits},
\author[fnal]{C.~James},
\author[ilit]{P.~A.~Kasper},
\author[fnal]{S.~Kwan},
\author[prin]{D.~C.~Langs},
\author[ucsc]{J.~Leslie},
\author[fnal]{B.~Lundberg},
\author[taun]{S.~MayTal-Beck},
\author[ucin]{B.~Meadows},
\author[cbpf]{J.~R.~T.~de~Mello~Neto},
\author[tuft]{R.~H.~Milburn},
\author[cbpf]{J.~M.~de~Miranda},
\author[tuft]{A.~Napier},
\author[ksun]{A.~Nguyen},
\author[ucin,pueb]{A.~B.~d'Oliveira},
\author[ucsc]{K.~O'Shaughnessy},
\author[ilit]{K.~C.~Peng},
\author[ucin]{L.~P.~Perera},
\author[usca]{M.~V.~Purohit},
\author[umis]{B.~Quinn},
\author[uwsc]{S.~Radeztsky},
\author[umis]{A.~Rafatian},
\author[ksun]{N.~W.~Reay},
\author[umis]{J.~J.~Reidy},
\author[cbpf]{A.~C.~dos Reis},
\author[ilit]{H.~A.~Rubin},
\author[umis]{D.~A.~Sanders},
\author[ucin]{A.~K.~S.~Santha},
\author[cbpf]{A.~F.~S.~Santoro},
\author[prin]{A.~J.~Schwartz},
\author[cine,uwsc]{M.~Sheaff},
\author[ksun]{R.~A.~Sidwell},
\author[yale]{A.~J.~Slaughter},
\author[ucin]{M.~D.~Sokoloff},
\author[cbpf]{J.~Solano},
\author[ksun]{N.~R.~Stanton},
\author[uwsc]{K.~Stenson},  
\author[umis]{D.~J.~Summers},
\author[yale]{S.~Takach},
\author[fnal]{K.~Thorne},
\author[osun]{A.~K.~Tripathi},
\author[uwsc]{S.~Watanabe},
\author[taun]{R.~Weiss-Babai},
\author[prin]{J.~Wiener},
\author[ksun]{N.~Witchey},
\author[yale]{E.~Wolin},
\author[umis]{D.~Yi},
\author[ksun]{S.~Yoshida},
\author[stan]{R.~Zaliznyak}, and
\author[ksun]{C.~Zhang}

\address[cbpf]{Centro Brasileiro de Pesquisas F{\'\i}sicas, Rio de Janeiro, Brazil}
\address[ucsc]{University of California, Santa Cruz, California 95064}
\address[ucin]{University of Cincinnati, Cincinnati, Ohio 45221}
\address[cine]{CINVESTAV, Mexico}
\address[fnal]{Fermilab, Batavia, Illinois 60510}
\address[ilit]{Illinois Institute of Technology, Chicago, Illinois 60616}
\address[ksun]{Kansas State University, Manhattan, Kansas 66506}
\address[umas]{University of Massachusetts, Amherst, Massachusetts 01003}
\address[umis]{University of Mississippi, University, Mississippi 38677}
\address[osun]{The Ohio State University, Columbus, Ohio 43210}
\address[prin]{Princeton University, Princeton, New Jersey 08544}
\address[pueb]{Universidad Autonoma de Puebla, Mexico}
\address[usca]{University of South Carolina, Columbia, South Carolina 29208}
\address[stan]{Stanford University, Stanford, California 94305}
\address[taun]{Tel Aviv University, Tel Aviv, Israel}
\address[stev]{Box 1290, Enderby, BC, V0E 1V0, Canada}
\address[tuft]{Tufts University, Medford, Massachusetts 02155}
\address[uwsc]{University of Wisconsin, Madison, Wisconsin 53706}
\address[yale]{Yale University, New Haven, Connecticut 06511}

\begin{abstract}
Using the large hadroproduced charm sample collected in experiment E791
at Fermilab, we have measured ratios of branching fractions
for the two-body singly-Cabibbo-suppressed charged decays of the $D^0$: 
$\kkovkpi   = 0.109 \pm 0.003 \pm 0.003$,
$\pipiovkpi = 0.040 \pm 0.002 \pm 0.003$, and
$\kkovpipi  = 2.75  \pm 0.15  \pm 0.16$. 
We have looked for differences in the decay rates of $D^0$ and $\overline 
D{}^0$ to the {\em CP} eigenstates $K^+K^-$ and $\pi^+\pi^-$, and have 
measured the {\em CP} asymmetry parameters 
$A_{CP}(K^+K^-)     = -0.010 \pm 0.049 \pm 0.012$ and 
$A_{CP}(\pi^+\pi^-) = -0.049 \pm 0.078 \pm 0.030$, 
both consistent with zero.

\end{abstract}
\

\end{frontmatter}
 
The measured world average for the ratio of the branching fractions of the
two-body singly-Cabibbo-suppressed (SCS) charged decays of the $D^0$ meson is
$\kkovpipi = 2.86 \pm 0.28$ \cite{pdg}.  Models including final state
interactions \cite{Kamal,Chau,bucc}, penguin diagrams \cite{Gluck}, QCD sum
rules \cite{Shif}, and non-perturbative algebraic approaches \cite{Terasaki} have
been proposed to explain the experimental value observed for this ratio.
Precise measurements of this ratio can help to differentiate among models, and
can also aid in our understanding of the Standard Model  predictions for
$D^0$--$\overline{D}{}^0$ mixing via long-range mechanisms, which require
$SU(3)$ symmetry-breaking to be non-zero \cite{Dono,Wolf,Kaed}. 

To date {\em CP} violation has been observed only in the neutral kaon system.
In the Standard Model this violation is a consequence of a complex amplitude in
the Cabibbo-Kobayashi-Maskawa matrix.  In this model, strange quarks couple to
the top quark in diagrams with internal loops, leading to observable {\em CP}
violation.  The comparable diagrams in the charm sector have bottom quarks in
the internal loops, resulting in very small Standard Model contributions to
{\em CP} violation.  Thus the charm sector may be uniquely sensitive to physics
outside the Standard Model, at the $10^{-3}$ level \cite{Xing}.  Currently, the
measured limits for charged and neutral $D$'s are at the level of (5--10)\%
from experiments E687 \cite{e6871,e6872} and E791 \cite{nader} at Fermilab, and
CLEO \cite{cleo90,cleo93}.

In this paper, we first report measurements of the branching fractions for the
SCS decays $D^0 \rightarrow K^+K^-$ and $D^0 \rightarrow \pi^+ \pi^-$  relative
to the Cabibbo-favored decay $D^0 \rightarrow K^-\pi^+$, based on a total 
sample of about 42,000 fully reconstructed two-body $D^0$ decays. Second, we
present a search for {\em CP} violation in the SCS decay of the $D^0$ using a
sample of 14,225 $D^0$'s tagged through the decay $D^{*+} \rightarrow
D^0\pi^+$. Throughout this paper, the {\em CP}-conjugate states are implicitly
included unless otherwise noted.  At the level of sensitivity of this
experiment, any direct {\em CP} violation would indicate physics outside the
Standard Model.

The current results are based on data accumulated by experiment E791 in a 500
GeV $\pi^-$ beam during the 1991/92 Fermilab fixed-target run. E791 was the
fourth in a series of charm experiments performed in the Fermilab Tagged Photon
Laboratory. The E791 spectrometer  \cite{exp:appel} was an open geometry
detector with 23 planes of silicon microstrip detectors (6 upstream and 17
downstream of the target), 35 drift chamber planes, 10 proportional wire
chambers (8 upstream and 2 downstream of the target), two magnets for momentum
analysis, two large multicell threshold \v{C}erenkov counters for charged
particle identification, electromagnetic and hadronic calorimeters for
electron/hadron separation as well as for online triggering, and a fast data
acquisition system that allowed us to collect data at a rate of 30 Mbyte/s with
a 50~$\mu$s/event deadtime. The target consisted of a 0.6-mm thick platinum
foil followed by four 1.5-mm diamond foils. Each target center was separated
from the next by about 1.5 cm, allowing observation of charmed-particle decays
in air without background from secondary interactions. The very open
transverse-energy trigger was based on the energy deposited in the calorimeters
and was highly efficient for charm events. Over $2 \times 10^{10}$ events were
recorded during a six-month period.

The two-body decay sample was selected based upon several criteria.  A decay
track candidate must have made a large contribution to the $\chi^2$ of the
event production vertex fit when included in that fit.  The significance of the
measured separation (in the beam direction) of the candidate decay vertex from
the production vertex had to be $ > 8 \sigma$, where $\sigma$ is the error on
the measured separation of the two points.  The momentum component of the $D^0$
candidate transverse to the line connecting the production and decay vertices
had to be less than $0.40\ {\rm GeV}/c$. The sum of $p_t^2$ of the decay
tracks, with $p_t$ measured relative to the direction of the $D^0$ candidate,
was required to be greater than $0.52\ ({\rm GeV}/c)^2$. Finally, the decay
vertex had to be located well outside the target foils.

\begin{figure}[htb]
\centerline {\epsfysize=4.5in \epsfxsize=6.5in \epsffile{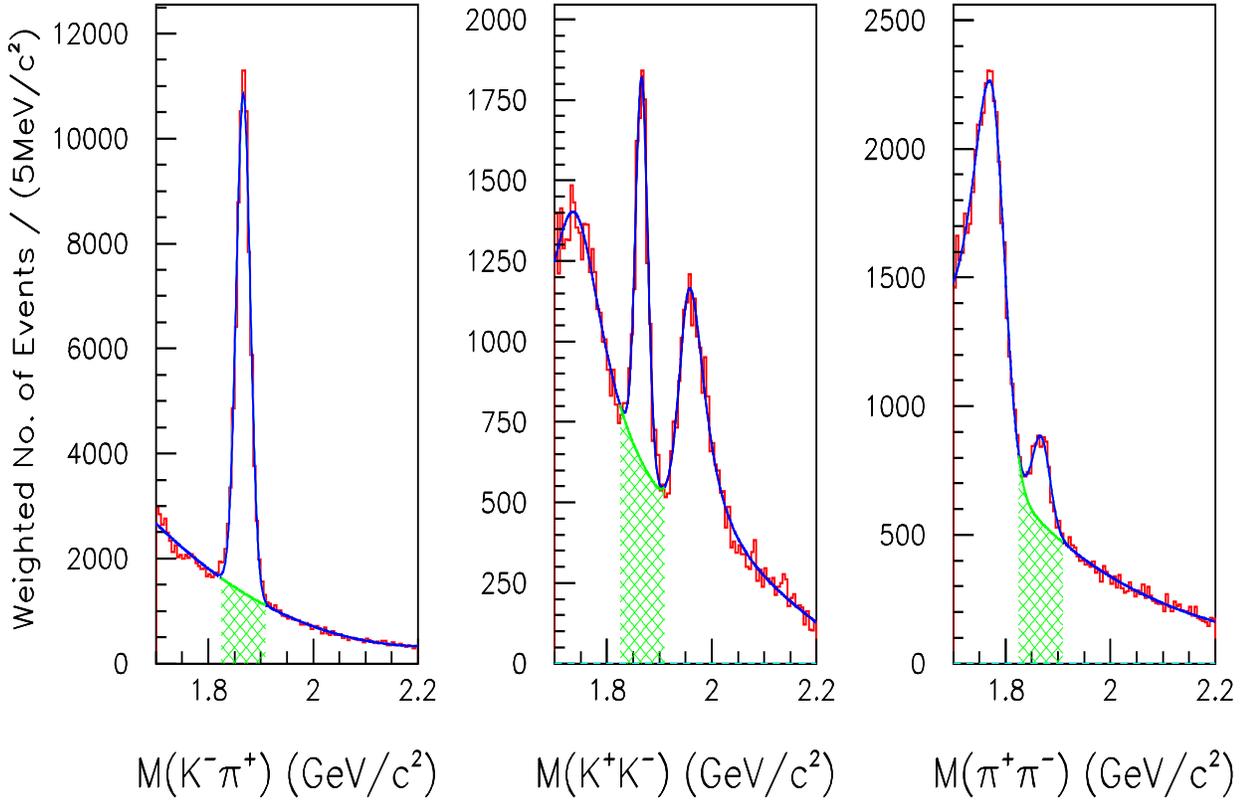} }
\caption{Invariant mass of \v{C}erenkov-weighted candidates. The solid lines 
         correspond to a binned maximum-likelihood fit to a signal given by a 
         Gaussian distribution plus different background 
         functions given by: $D^0 \rightarrow K^-\pi^+$ -- a third-order 
         polynomial (left); $D^0 \rightarrow K^+K^-$ -- a linear term plus a 
         Breit-Wigner function for misidentified $K\pi\pi^0$ and a
         half-Gaussian-half-Breit-Wigner function for misidentified $K\pi$ 
         (middle); and $D^0 \rightarrow \pi^+\pi^-$ -- an exponential term 
         plus a half-Breit-Wigner-half-Gaussian function for misidentified 
         $K\pi$ (right). Cross-hatched areas are our estimates of 
         backgrounds below the signals.}
\label{fg:br}
\end{figure}

To improve the statistical significance of the ratio $\kkovpipi$, we required 
the daughter tracks to have $K$ or $\pi$ signatures in  the threshold
\v{C}erenkov counters. However, the efficiency of our identification depends on
particle momentum. We have studied these particle identification efficiencies
for the data sample $D^0 \rightarrow K^- \pi^+$ as a function of momentum,
transverse momentum with respect to the beam direction, charge, and particle
type. Based on this study, we required the particle momenta to be in the range
6 to 80 ${\rm GeV}/c$. We then corrected for the particle identification
efficiency by weighting each $D^0$ candidate by the inverse of the product of
the two particle identification efficiencies.  Using this procedure we have
weighted each track according to its $p$ and $p_t$, rather than applying one
global factor to the final results.  We calculated the statistical errors in
the weighted signals by scaling the statistical errors of the unweighted
signals by the ratios of the weighted to unweighted sample sizes. 
Figure \ref{fg:br} shows the \v{C}erenkov-weighted
invariant mass distributions for our data. The peaks to the right of the signal
region in the $K^+K^-$ distribution and to the left of the signal region in the
$\pi^+\pi^-$ distribution are due to misidentified $K\pi$ events, whereas the
peak to the left of the signal region in the $K^+K^-$  distribution is due to
misidentified $K\pi\pi^0$ events.  A binned maximum-likelihood fit was
performed for each distribution with the signal function assumed to be
Gaussian.  The backgrounds were fit to the following functions: $K^-\pi^+$ -- a
third-order polynomial (Figure \ref{fg:br} left); $K^+K^-$ -- a linear term
plus a  Breit-Wigner function for misidentified $K\pi\pi^0$ and a
half-Gaussian-half-Breit-Wigner function for misidentified $K\pi$ (Figure
\ref{fg:br} middle); $\pi^+\pi^-$ -- an exponential term plus a
half-Breit-Wigner-half-Gaussian for misidentified $K\pi$ (Figure \ref{fg:br}
right). Cross-hatched areas are the estimated backgrounds under the signals.

We have used a Pythia-based Monte Carlo (MC) \cite{sjo} which incorporates a
full detector simulation to correct for detector acceptance effects other than
\v{C}erenkov identification. The efficiencies determined from the Monte Carlo
simulation and the number of signal events extracted from both the unweighted
and weighted mass distributions are given in Table~\ref{tb:br}. 

\begin{table}[ht]
\caption{Sample sizes and average geometric acceptances for three decay modes 
         of the $D^0$.
\label{tb:br}}
\tabular{cccc}
\hline
  & $D^0 \rightarrow K^-\pi^+$ & 
    $D^0 \rightarrow K^+K^-$   & $D^0 \rightarrow \pi^+\pi^-$ \\
\hline
Unweighted Sample   & $36955 \pm 217$ &  $3317 \pm \phantom{1}84$ & $2043 \pm
\phantom{1}95$ \\
Weighted Sample     & $63177 \pm 371$ &  $6845 \pm 172$ & $2521 \pm 117$ \\
Average Acceptance  & 3.21\%          &  3.18\%         & 3.22\% \\ 
\hline
\endtabular
\end{table}

We calculate the ratios of branching fractions from the numbers in the last two
rows of Table~\ref{tb:br} and present them together with previous measurements
in Table~\ref{tb:combr}. The quoted systematic errors are the quadrature sums
of systematic uncertainties from the following sources: relative efficiencies
for the selection criteria, fitting functions, MC production models, and the
\v{C}erenkov particle identification weighting procedure. Table~\ref{tb:brsyst}
presents the contribution of each of these to the total systematic uncertainty,
expressed as a percentage of the statistical uncertainty.  The ``Selection
Criteria'' entry reflects uncertainties in the details of the MC modeling of
the experimental acceptance.  The ``Fitting Functions'' uncertainty corresponds
to various estimates of the background, especially for $\pi\pi$.  We note that
$K\pi$ reflections do not contribute directly in the $KK$ or $\pi\pi$ signal
regions.  The shapes of the $K\pi$ and $K\pi\pi^0$ reflections have been
studied in detail, and the same fitting function yields good fits to both MC
and data. Uncertainties resulting from the MC production model were
investigated  by changing PYTHIA default parameters to agree with a study of
$D^\pm$ production in E791\cite{carter}.

\begin{table}[ht]
\caption{$D^0 \rightarrow K^+K^-$ and $\pi^+\pi^-$ relative branching ratio 
         measurements compared with previous experiments.\label{tb:combr}}
\tabular{ccl@{${}\pm{}$}l@{${}\pm{}$}ll@{${}\pm{}$}l@{${}\pm{}$}ll
         @{${}\pm{}$}l@{${}\pm{}$}l}
\hline
Year & Group & \multicolumn{3}{c}{\kkovkpi} &
   \multicolumn{3}{c}{\pipiovkpi} & \multicolumn{3}{c}{\kkovpipi} \\
\hline
1997 & E791 & 
	0.109	& 0.003 & 0.003	&
	0.040	& 0.002 & 0.003	&
	2.75	& 0.15	& 0.16	\\
\hline \hline
1979 & Mark II \cite{mkii} &
       0.113 & \multicolumn{2}{@{}l}{0.030} &
       0.033 & \multicolumn{2}{@{}l}{0.015} &
         3.4 & \multicolumn{2}{@{}l}{1.8} \\
1984 & Mark III \cite{mkiii} &
       0.122 & 0.018 & 0.012 &
       0.033 & 0.010 & 0.006 &
         3.7 & \multicolumn{2}{@{}l}{1.3} \\
1989 & ARGUS \cite{argus} &
       0.10  & 0.02  & 0.01  &
       0.040 & 0.007 & 0.006 &
         2.5 & \multicolumn{2}{@{}l}{0.7} \\
1990 & CLEO \cite{cleo90}  &
       0.117 & 0.010 & 0.007 &
       0.050 & 0.007 & 0.005 &
        2.35 & 0.37  & 0.28  \\
1991 & E691 \cite{e691}   &
       0.107 & 0.010 & 0.009 &
       0.055 & 0.008 & 0.005 &
        1.95 & 0.34  & 0.22  \\
1992 & WA82 \cite{wa82} &
       0.107 & 0.029 & 0.015 &
       0.048 & 0.013 & 0.008 &
       2.23  & 0.81  & 0.46  \\
1993 & CLEO \cite{cleo93} &
       \multicolumn{3}{c}{} &
       0.0348 & 0.0030 & 0.0023 &
       \multicolumn{3}{c}{} \\
1994 & E687 \cite{e6871} &
       0.109 & 0.007 & 0.009 &
       0.043 & 0.007 & 0.003 &
       2.53  & 0.46  & 0.19  \\
1996 & PDG \cite{pdg} &
       0.113 & \multicolumn{2}{@{}l}{0.006} &
       0.0396 & \multicolumn{2}{@{}l}{0.0027} &
       2.86  & \multicolumn{2}{@{}l}{ 0.28} \\
\hline
\endtabular
\end{table}
\begin{table}[ht]
\caption{Contributions to the systematic uncertainty for each of the measured
         branching ratios. The systematic uncertainties are expressed as a 
         percentage of the statistical uncertainty for the corresponding 
         branching ratio.
\label{tb:brsyst}} 
\tabular{lccc}
\hline
  & $\kkovkpi$ & $\pipiovkpi$ & $\kkovpipi$ \\
\hline
Selection Criteria    &  90\% & 100\% &  40\% \\
Fitting Functions     &  60\% &  70\% &  90\% \\
MC Production Model   &  30\% &  30\% &  20\% \\
Particle ID Weighting &  30\% &  30\% &  30\% \\
\hline
Total                 & 116\% & 129\% & 105\% \\
\hline
\endtabular
\end{table}

We now present results of a search for {\em CP} violation in the SCS decays of
the $D^0$ (or $\overline{D}{}^0$) beginning with the same data sample as
above.  Assuming {\em CP} conservation in the strong decay of the $D^*$, we
identify the meson as either $D^0$ or $\overline{D}{}^0$ by tagging it using
the charge of the slow $\pi$ from $D^{*+} \rightarrow D^0 \pi^+$ and $D^{*-}
\rightarrow \overline{D}{}^0 \pi^-$.  We required the mass difference between
$D^*$ and $D$ to be in the range 143 -- 148 ${\rm MeV}/c^2$ and the distance of
closest approach of the bachelor pion to the primary vertex had to be $<
120~\mu$m. Because $D^*$ tagging reduces background so strongly, we relaxed the
requirement on the sum of $p_t^2$ of the $D^0$ decay tracks relative to the
direction of the parent $D^0$ from $0.52~{\rm (GeV}/c)^2$ to $0.4~{\rm
(GeV}/c)^2$. For this analysis, we used the same \v{C}erenkov identification
criteria as for the branching ratio measurements, but we did not weight the
events.

The signal for {\em CP} violation is an absolute rate difference between decays
of particle and antiparticle to charge-conjugate final states $f$ and
$\overline f$, where $f = K^+K^-$ or $\pi^+\pi^-$: 
\begin{equation}  
A_{CP} = \frac{\Gamma(D\rightarrow f)-\Gamma(\overline{D}\rightarrow\overline f)}
{\Gamma(D\rightarrow f)+\Gamma(\overline{D}\rightarrow\overline f)}. 
\end{equation}  

In hadroproduction, $D$ and $\overline{D}$ mesons are not produced equally.
Therefore we normalized our signals to the Cabibbo-favored
$D^0 \rightarrow K^- \pi^+$ and $\overline{D}{}^0 \rightarrow K^+ \pi^-$ signals 
and measured
\begin{equation}
\frac {\frac{\displaystyle{N(D^0\rightarrow f)}} 
            {\displaystyle{N(D^0\rightarrow K^-\pi^+)}} - 
       \frac{\displaystyle{N(\overline{D}{}^0\rightarrow \overline f)}}
            {\displaystyle{N(\overline{D}{}^0\rightarrow K^+\pi^-})}}
      {\frac{\displaystyle{N(D^0\rightarrow f)}} 
            {\displaystyle{{N(D^0\rightarrow K^-\pi^+)}}} + 
       \frac{\displaystyle{N(\overline{D}{}^0\rightarrow \overline f)}}
            {\displaystyle{N(\overline{D}{}^0\rightarrow K^+\pi^-})}} 
\end{equation} 
where, for each channel, $N = \epsilon n$ is the observed number of events,
$n$ is the produced number of events, and $\epsilon$ is the efficiency of our
detector and analysis procedure.  To the extent that 
\begin{equation} 
\frac {\epsilon(D^0 \rightarrow f)}{\epsilon(D^0 \rightarrow K^-\pi^+)} =  
\frac {\epsilon(\overline{D}{}^0 \rightarrow \overline f)}
      {\epsilon(\overline{D}{}^0 \rightarrow K^+\pi^-)}, 
\end{equation}
the measured quantity (Eq.~2) is $A_{CP}$. We verified the validity of Eq.~3
for this analysis in two stages. First, we used real data to determine that the
\v{C}erenkov identification efficiencies for kaons and pions are independent of
charge in any given momentum range from 6 to 80 ${\rm GeV}/c$.  Then, we used
our Monte Carlo simulation of the experiment to determine that the geometric
acceptances are independent of charge.  

Note that any {\em CP} asymmetry from interference between mixing and tree-level
diagrams will not cancel through the $D^0 \rightarrow K^- \pi^+$ normalization.
Thus our normalized $A_{CP}$ is not a direct {\em CP} asymmetry parameter, but
rather a measure of combined direct and indirect {\em CP}
asymmetries \cite{PaWu}. An implicit assumption in this analysis is that there
is no measurable {\em CP} violation in the Cabibbo-favored decays of the $D^0$.

\begin{figure}[htb]
\centerline {\epsfxsize=6.5in \epsffile{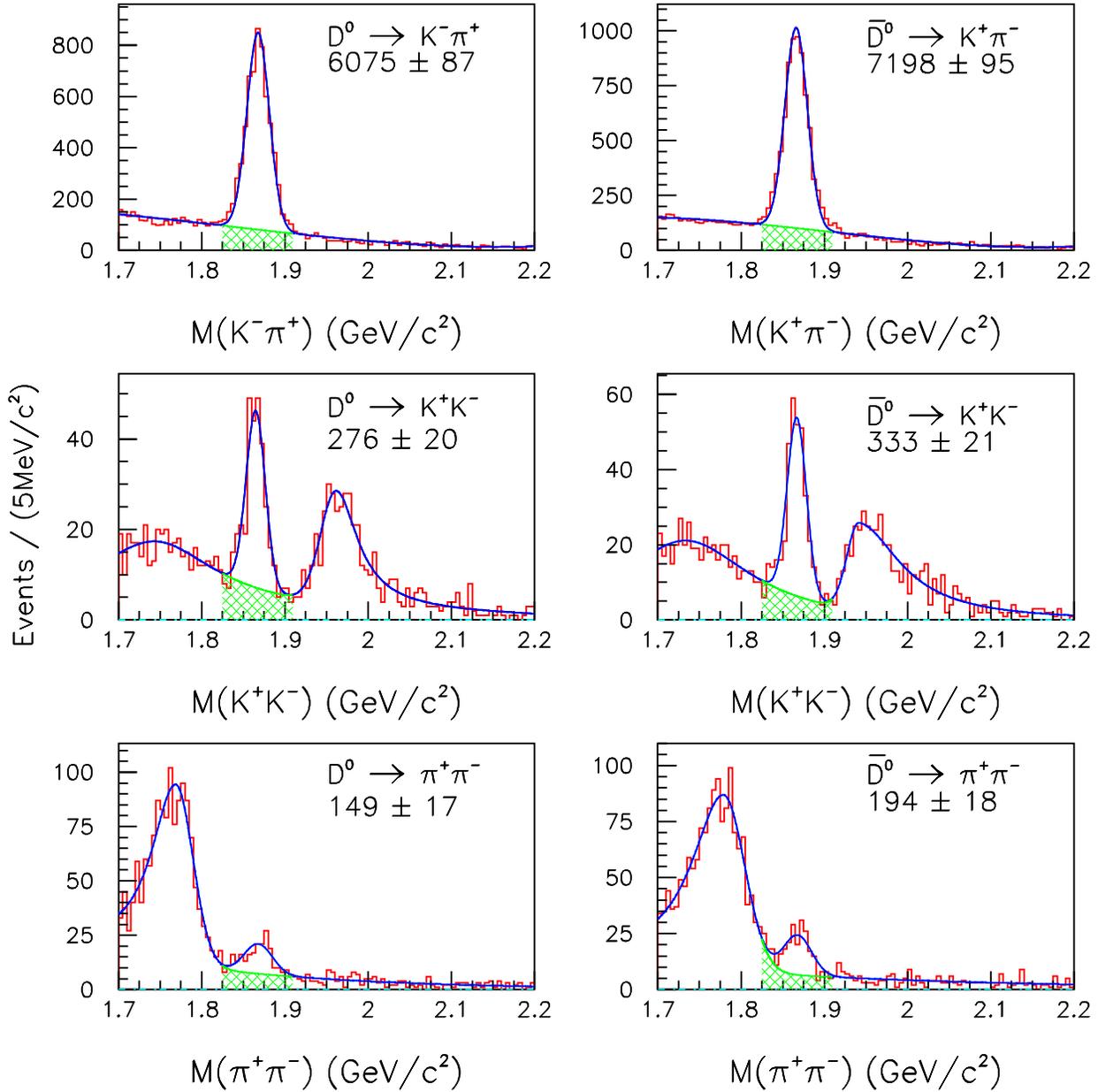} }
\vspace{0.15in}
\caption{Invariant mass of $D^0$ candidates (left column of figures) and 
         $\overline {D}{}^0$ (right column of figures) tagged with the decay 
         $D^* \rightarrow D^0\pi$, for the decay \kpi ~(top), \kk ~(middle), 
         and \pipi~(bottom). The numbers presented are the sample sizes as 
         calculated from Gaussian fits to the signal regions, and the
         cross-hatched areas are our estimates of backgrounds below the 
         signals.}
\label{fg:cp}
\end{figure}

Figure \ref{fg:cp} presents mass plots for the candidate $D^0$ and $\overline
D{}^0$ decays to $K\pi$, $KK$, and $\pi\pi$, including our fits to the
distributions.  We used the same fixed central mass value and signal width for
$D^0$ and $\overline{D}{}^0$ fits. The distributions fit the data well, and
the integrals of the Gaussian fits to the signals were used to calculate the
observed numbers of decays.  Our asymmetry results are listed in
Table~\ref{tb:comacp} along with previous measurements. We see no evidence of
{\em CP} violation.
\begin{table}[ht]
\caption{The measured asymmetry parameters $A_{CP}$ for $K^+K^-$ and 
         $\pi^+\pi^-$ decay modes of the $D^0$ and their 90\% confidence-level 
         limits, compared with previous experiments.
\label{tb:comacp}}
\tabular{ccr@{${}\pm{}$}l@{${}\pm{}$}lr@{${}<{}$}c@{${}<{}$}r}
\hline
Year & Group & \multicolumn{3}{c}{$A_{CP}$} & 
               \multicolumn{3}{c}{Limit at 90\% confidence level} \\
\hline
1997 & E791 &
       $-0.010$ & 0.049 & 0.012 
       & $-9.3\%$ & $A_{CP}(K^+K^-)$ & 7.3\% \\
1997 & E791 &
       $-0.049$ & 0.078 & 0.030 
       & $-18.6\%$ & $A_{CP}(\pi^+\pi^-)$ & 8.8\% \\
\hline \hline
1991 & E691 \cite{e691} &
       0.20\phantom{4}  &  \multicolumn{2}{@{}l}{0.15 } 
       & \multicolumn{1}{r}{}&{$A_{CP}(K^+K^-$)} & 45\% \\
1994 & E687 \cite{e6872} &
       0.024 & \multicolumn{2}{@{}l}{0.084} 
       & $-11\%$  & $A_{CP}(K^+K^-)$ & 16\% \\
1995 & CLEO II \cite{cleo95} &
       0.080 & \multicolumn{2}{@{}l}{0.061} 
       & $-2.2\%$ & $A_{CP}(K^+K^-)$ & 18\% \\
\hline
\endtabular
\end{table}

In summary, we have measured the following ratios of branching fractions for
the charged two-body decays of the $D^0$: $\kkovkpi = 0.109 \pm 0.003 \pm
0.003$, $\pipiovkpi = 0.040 \pm 0.002 \pm 0.003$, and $\kkovpipi = 2.75 \pm
0.15 \pm 0.16$.  We also find the 90$\%$ confidence-level intervals on the
decay asymmetries of $D^0$ and $\overline{D}{}^0$ to the {\em CP} eigenstates
$K^+K^-$ and $\pi^+\pi^-$ to be $-9.3\% < A_{CP}(K^+K^-) < 7.3\%$ and $-18.6\%
< A_{CP}(\pi^+\pi^-) < 8.8\%$. We find no evidence for {\em CP} violation in
these decay modes. This is the first reported result of a search for {\em CP}
violation in the decay mode $D^0 \rightarrow \pi^+\pi^-$.

We gratefully acknowledge the assistance of the staffs of Fermilab and
of all the participating institutions. This research was supported by
the Brazilian Conselho Nacional de Desenvolvimento Cient\'\i fico e
Tecnol\'ogico, CONACyT (Mexico), the Israeli Academy of Sciences and
Humanities, the U.S. Department of Energy, the U.S.-Israel Binational
Science Foundation, and the U.S. National Science Foundation. Fermilab
is operated by the Universities Research Association, Inc., under
contract with the United States Department of Energy.

\bibliographystyle{unsrt}

\begin{thebibliography}{99}
\bibitem {pdg} Particle Data Group, Phys.\ Rev.\ D {\bf 54} (1996) 33.
\bibitem {Kamal} A. N.\ Kamal {\em et al.}, Phys.\ Rev.\ D {\bf 35} (1987) 3515;
A. N. Kamal {\em et al.}, Phys.\ Rev.\ D {\bf 36} (1987) 3510;
A. N. Kamal {\em et al.}, Z. Phys.\ C {\bf 54} (1992) 411;
A. N. Kamal {\em et al.}, Phys.\ Rev.\ D {\bf 50} (1994) 1832.
\bibitem{Chau} L. L. Chau {\em et al.}, Phys.\ Lett.\ {\bf B280} (1992) 281;
L. L.\ Chau {\em et al.}, Phys.\ Lett.\ {\bf B333} (1994) 514.
\bibitem {bucc}  F. Buccella, {\em et al.}, Phys.\ Rev.\ D {\bf 51} (1995) 3478;
F. Buccella, M. Lusignoli, and A. Pugliese, Phys.\ Lett.\ {\bf B379} (1996) 249.
\bibitem {Gluck} M. Gl\"{u}ck, Phys.\ Lett.\ {\bf B88} (1979) 145.
\bibitem {Shif} B. Yu.\ Block and M. A.\ Shifman, Sov.\ J.\ Nucl.\ Phys. 
{\bf 45} (1987) 135; {\em ibid.} {\bf 45} (1987) 301; {\em ibid.} 
{\bf 45} (1987) 522. 
\bibitem {Terasaki} K. Terasaki {\em et al.}, Phys.\ Rev.\ D {\bf 38} (1988) 
132.
\bibitem {Dono} J. F. Donoghue, E. Golowich, B. R. Holstein, and J. Trampetic, 
Phys.\ Rev.\ D {\bf 33} (1986) 179.
\bibitem {Wolf} L. Wolfenstein, Phys.\ Lett.\ B {\bf 164} (1985) 170.
\bibitem {Kaed} T. A. Kaeding, Phys.\ Lett.\ B {\bf 357} (1995) 151.
\bibitem {Xing} Z. Xing, Phys.\ Rev.\ D {\bf 55} (1997) 196, and references
therein.
\bibitem {e6871} P. L. Frabetti {\em et al.}, Phys.\ Lett.\ {\bf B321} (1994) 
295.
\bibitem {e6872} P. L. Frabetti {\em et al.}, Phys.\ Rev.\ D {\bf 50} (1994) 
2953.
\bibitem {nader} E. M. Aitala {\em et al.}, Phys.\ Lett.\ {\bf B403} (1997) 377.
\bibitem {cleo90} J. Alexander {\em et al.}, Phys.\ Rev.\ Lett.\ {\bf 65} (1990)
1184.
\bibitem {cleo93} M. Selen {\em et al.}, Phys.\ Rev.\ Lett.\ {\bf 73} (1993) 
1973.
\bibitem {exp:appel} J. A. Appel, Ann.\ Rev.\ Nucl.\ Part.\ Sci.\ {\bf 42}
(1992) 367, and references therein; 
D. J. Summers {\em et al.}, Proceedings of the {\it XXVII$^{\,th}$ Rencontre de 
Moriond}, Electroweak Interactions and Unified Theories, Les Arcs, France 
(15-22 March, 1992) 417;
S. Amato {\em et al.}, Nucl.\ Instr.\ Meth.\ {\bf A324} (1993) 535; S. Bracker 
{\em et al.}, IEEE Trans.\ Nucl.\ Sci.\ {\bf NS-43} (1996) 2457; E.
M. Aitala {\em et al.}, Phys.\ Rev.\ Lett.\ {\bf 76} (1996) 364.
\bibitem {sjo} H.-U. Bengtsson and T. Sj\"{o}strand, Comp.\ Phys.\ Comm.\ 
{\bf 46} (1987) 43.
\bibitem {carter} E. M. Aitala {\em et al.}, Phys.\ Lett.\ {\bf B371} (1996) 
157. 
\bibitem {PaWu} W. F. Palmer and Y. L. Wu, Phys.\ Lett.\ {\bf B350} (1995) 245.
\bibitem {mkii} G. S. Abrams {\em et al.}, Phys.\ Rev.\ Lett.\ {\bf 43} (1979) 
481.
\bibitem {mkiii} R. M. Baltrusaitis {\em et al.}, Phys.\ Rev.\ Lett.\ {\bf 55}
(1985) 150.
\bibitem {argus} H. Albrecht {\em et al.}, Zeit. Phys. C {\bf 46} (1990) 9.
\bibitem {e691} J. C. Anjos {\em et al.}, Phys.\ Rev.\ D {\bf 44} (1991) R3371.
\bibitem {wa82} M. Adamovich {\em et al.}, Phys.\ Lett.\ {\bf B280} (1992) 163.
\bibitem {cleo95} J. Bartelt {\em et al.}, Phys.\ Rev.\ D {\bf 52} (1995) 4860.
 
\end{thebibliography}

\end{document}